\documentclass[twocolumn,preprintnumbers,
nofootinbib,prd]{revtex4}

\usepackage{graphicx}

\usepackage[hypertex]{hyperref}
\newcommand{\vev}[1]{\langle {#1} \rangle}
\newcommand{\eq}[1]{Eq.~(\ref{#1})}
\newcommand{\lsim}{\lesssim}

\newcommand{\gsim}{\gtrsim}
\newcommand{\beq}{\begin{equation}}
\newcommand{\eeq}{\end{equation}}
\newcommand{\LT}{\Lambda_T}
\newcommand{\lqcd}{\Lambda_{\rm QCD}}
\newcommand{\qq}{\vev{{\bar Q_L}{Q}_R}}

\begin{document}

\pagestyle{plain}

\preprint{MADPH-06-1468}

\title{Cosmology with Light Axions from Technicolor}

\author{Hooman Davoudiasl\footnote{\tt email: hooman@physics.wisc.edu}}

\affiliation{Department of Physics, University of Wisconsin,
Madison, WI 53706, USA}


\begin{abstract}

We consider cosmological consequences of the spontaneous breaking of a global symmetry that is anomalous under technicolor interactions, leading to the emergence of a light axion-like particle.  Avoiding overclosure of the
universe by such axions yields the upper bound $f_a \lsim 10^{10}$~GeV
on the symmetry breaking scale, corresponding to keV-scale axions.  However, diffuse x-ray background data typically require larger values of $f_a$.  The overclosure and x-ray bounds can be reconciled if the axion initial amplitude of oscillations $A_i \sim f_a/10$.  In this case, a viable axionic dark matter candidate with a mass in the $50-100$~eV range emerges.  The detection of this type of dark matter may pose a challenge.

\end{abstract}
\maketitle


A main goal of future experiments at high energies is to uncover
the mechanism for electroweak symmetry breaking (EWSB).  The most
economic proposal employs the Higgs doublet of the Standard Model
(SM). However, this simple picture is unstable against quantum
corrections.

There are only a few theoretical frameworks for stable EWSB. One
such framework postulates the existence of a new interaction that
results in EWSB via confinement near the weak scale $M_W \sim
100$~GeV\cite{Weinberg:1975gm}. This mechanism is a higher energy
analogue of QCD and, therefore, often called Technicolor. Here,
the smallness of the weak scale compared to, say, the gravity
scale $M_P \sim 10^{19}$~GeV, is a consequence of dynamics, like
the proton mass in QCD, and no longer a puzzle. Since QCD is the
only known mechanism for generation of microscopic mass scales in
Nature, technicolor is a well-motivated construct.

Confinement in QCD can have other interesting consequences besides generating the hadronic scale $\Lambda_{\rm QCD} \sim
200$~MeV.  For example, if a spontaneously broken global symmetry is anomalous under QCD, as in the Peccei-Quinn (PQ) mechanism \cite{Peccei:1977hh}, the Goldstone boson, generally called an axion, associated with the broken symmetry \cite{Weinberg:1977ma} acquires a mass through confinement at $\Lambda_{\rm QCD}$.  The PQ mechanism was originally devised to explain the tiny size of the CP violating angle $\theta_{\rm QCD} < 10^{-10}$ in QCD \cite{Peccei:1977hh}, as suggested by data. QCD dynamics generates a mass $m \sim
\lqcd^2/f$ for the axion, where $f$ is the scale of PQ symmetry breaking.  
The coupling of the axion to the SM is
set by $1/f$.  Astrophysical data suggest $f \gsim 10^9$~GeV
\cite{Eidelman:2004wy}.
Hence, the PQ axion must be very light and extremely weakly
coupled. 

Remarkably, cosmological overclosure arguments also yield
an upper bound $f \lsim 10^{12}$~GeV 
\cite{Preskill:1982cy,Abbott:1982af,Dine:1982ah};
near the upper bound, the axion can be
a good dark matter candidate.  In this paper, we consider the analogous question within a technicolor-like framework.  That is, we investigate the cosmological implications of breaking a techni-anomalous global symmetry (henceforth generically referred to as a PQ symmetry, but unrelated to the one relevant for $\theta_{\rm QCD}$), at a high scale $f_a$ \cite{TCcosmo}.  The resulting axion $a$ can pick up a mass $m_a \sim M_W^2/f_a$.  Therefore, $a$ can be very light and weakly coupled if $f_a \gg M_W$.

The problem of cosmological 
domain walls from axion dynamics \cite{Sikivie:1982qv}  
is not addressed here.  This problem can 
be avoided if, for example, 
inflation takes place after PQ symmetry breaking, 
rsulting in a constant axion background throughout the visible universe  
\cite{Preskill:1982cy,Abbott:1982af,Dine:1982ah,PS}. We also note that light particles that could in principle affect cosmology have been discussed in other contexts, such as supersymmetry \cite{GM}.  Our work suggests that technicolor can in principle be another context for a cosmologically important light particle.

Before we begin our analysis, we make a few remarks.  First of all, we do not address model-building issues related to a specific techni-anomalous PQ construct.  However, as PQ symmetry breaking ususally involves a non-zero scalar vacuum expectation value (vev) and technicolor is invoked to eliminate the need for scalars, one may object that our assumptions are conceptually inconsistent.  Here we note that this is not the case, since we will show that various considerations typically require $f_a$ to be in excess of $10^{10}$~GeV.  If the cutoff of a theory is near such scales, as may be the case with extra dimensions and the quantum gravity scale $M_F \gsim 10^{10}$~GeV, then PQ symmetry breaking near $M_F$ is quite natural.  In the absence of low energy supersymmetry, technicolor-like strong dynamics is still well-motivated in order to keep $M_W$ hierarchically below $f_a \sim M_F \gsim 10^{10}$~GeV.  Hence, our setup can be consistently embedded within a large class of models.

Next, we will
introduce the framework that will be used to study the effects of
technicolor strong dynamics and high scale PQ symmetry
breaking \footnote{We will only consider technicolor and ignore QCD. However,
the original strong CP problem may be addressed by postulating a
second PQ symmetry.}.  Later, we will consider this setup in the early
universe, at temperatures relevant to EWSB.
We will only focus on key theoretical
features of technicolor. Our conclusions will then be
applicable to a wide range of models with acceptable low energy
physics.  We find that the upper and lower bounds on $f_a$, from overclosure and diffuse x-ray background considerations, respectively, can be reconciled if the initial amplitude of primordial axion oscillations is an order of magnitude below $f_a$.  Hence, with a mild suppression of the initial amplitude, a dark matter candidate in the mass range $50-100$~eV can emerge.

We will assume that EWSB is realized by the condensation of a
chiral fermion bilinear $\qq \approx \LT^3$, with the correct
$SU(2)\times U(1)$ quantum numbers to result in the observed SM
pattern.  The condensation is caused by strong dynamics of a
technicolor gauge group $SU(N_T)$ at the scale $\LT \sim M_W$.
Further, we assume that $N_D$ left-handed weak doublet and $2 N_D$
right-handed weak singlet techni-quarks are in the $N_T$
fundamental representation.  Thus, the theory is endowed with a
$SU(2 N_D)_L\times SU(2 N_D)_R$ chiral symmetry. Upon EWSB, $(2
N_D)^2-1$ Pseudo-Goldstone Bosons (PGB's) are generated.  These
particles are the analogues of pions in QCD. We will assume
that the techni-sector PGB's have a decay constant $F_T$.

The qualitative and approximate quantitative features of
technicolor gauge interactions are deduced by direct analogy with
QCD, as strong dynamics is not theoretically
well-understood.  Since we are interested only in general
properties of technicolor, this approach has sufficient accuracy
for our purposes.  We will use the scaling rules presented in
Ref.~\cite{Hill:2002ap} in our work.

The scale at which strong dynamics sets in is given by
\beq
\LT \approx \lqcd (v_0/f_\pi)
\left(\frac{3}{N_D N_T}\right)^{1/2},
\label{LT}
\eeq
where $\lqcd = 200$~MeV, $v_0 = 246$~GeV
is the scale of EWSB in the SM, and
$f_\pi = 93$~MeV is the pion decay constant.
The PGB decay constant $F_T$ is given by
\beq
F_T \approx v_0 N_D^{-1/2}.
\label{FT}
\eeq

We assume that techni-quarks have a PQ chiral symmetry that is
spontaneously broken at a high scale $f_a \gg \LT$. The massless
axion corresponding to the broken symmetry will acquire a mass
$m_a$ due to the $SU(N_T)$ techni-anomaly.  We
will derive a low energy expression for $m_a$ 
by analogy with the results from QCD.

The QCD related axion mass is given by \cite{Georgi:1986df}
\beq
m_a^{(QCD)\,2} =
\frac{m_\pi^2 f_\pi^2}{(m_u + m_d)\,f^2\,
{\rm Tr}(M^{-1})},
\label{ma2qcd}
\eeq
where $M = {\rm diag}(m_u,m_d,m_s)$ is the diagonal light
quark mass matrix.  Let $m_0$ be the typical techni-pion mass,
generated by explicit chiral symmetry breaking.  In QCD, this is
due to the masses of the light quarks.  However, in technicolor,
there could be several sources of chiral symmetry breaking.  We
will parameterize this effect by assuming a common techni-quark
mass $m_Q$.  Even though this will not encode all aspects
of chiral symmetry breaking in technicolor, our approach will
capture the relevant physics and yield a
good estimate for the size of the effects.  Hence, in
Eq.~(\ref{ma2qcd}), we make the replacements $m_u + m_d \to 2 m_Q$,
$m_\pi \to m_0$, $f_\pi \to F_T$, and ${\rm Tr}(M^{-1}) \to
2 N_D/m_Q$.  Using \eq{FT}, we find
\beq
m_a \approx \frac{m_0 v_0}{2 N_D f_a} \quad
({\rm Technicolor}).
\label{maTC}
\eeq
Current experimental bounds require
$m_0 \gsim 100$~GeV \cite{Eidelman:2004wy}.

In order to study the cosmological implications of the above
setup, we will need to consider what happens to the axion when
strong techni-dynamics sets in around the weak-scale in the early
universe.  Generically, due to the techni-anomaly of the PQ
symmetry, the axion will be endowed with a potential and pick up a
temperature dependent mass $m_a(T)$.
Once $m_a(T)\simeq 3 H(T)$, where
$H$ is the Hubble constant, the axion will start to
oscillate with an amplitude $A$ \cite{Preskill:1982cy}.
Assuming that the zero-mode axion field does
not decay on the time scales we will consider, its energy gets
red-shifted according to \cite{Preskill:1982cy}
\beq
\frac{[m_a
A^2]_f}{[m_a A^2]_i} = \left(\frac{R_i}{R_f}\right)^3,
\label{redshift}
\eeq
where $R$ is the scale factor and the
subscripts $i$ and $f$ refer to initial and final values,
respectively.  In deriving \eq{redshift}, it has been assumed
that $m_a^{-1}(dm_a/dt), H \ll m_a$ \cite{Preskill:1982cy}.
Conservation of entropy in equilibrium then yields
\beq
[m_a A^2]_f = [m_a A^2]_i \left(\frac{g_{sf}}{g_{si}}\right)
\left(\frac{T_f}{T_i}\right)^3,
\label{entropy}
\eeq
where $g_s$ refers to the number of relativistic degrees of freedom
in equilibrium and $T_i$ is the temperature at which the axion
starts to oscillate.

The energy density stored in the axion zero mode is given by
$\rho_a = (1/2)\, m_a^2 A^2$.  Using \eq{entropy}, we can write
\beq
\rho_{a f} = \frac{1}{2}\, m_{a f} \,
m_{a i} A_i^2\left(\frac{g_{s f}}{g_{s i}}\right)
\left(\frac{T_f}{T_i}\right)^3.
\label{rhoaf}
\eeq
The Hubble scale during radiation domination is given by
$H = 1.66 \, g_i^{1/2} (T^2/M_P)$, with $M_P \simeq 1.2 \times
10^{19}$~GeV; $g_{si}=g_i$ for $T_i \gg m_e$.  We get
\beq
m_{ai}\equiv m_a(T_i) \simeq 5\, g_{si}^{1/2}
\frac{T_i^2}{M_P}.
\label{maTi}
\eeq
Eqs.~(\ref{rhoaf}) and (\ref{maTi}) together yield
\beq
\rho_a(T_f) =
\left(\frac{5\, g_{sf}}{2 \,g_{si}^{1/2}}\right)
\left(\frac{f_a}{M_P}\right)
\left(\frac{m_{af} f_a}{T_i}\right)
\left(\frac{A_i}{f_a}\right)^2 T_f^3.
\label{rhoaTf}
\eeq

The above equation can be used to obtain a bound on
the PQ scale $f_a$ by requiring that axions do not overclose
the universe \cite{Preskill:1982cy}.  Here, we implicitly assume that
the axion has a sufficiently long lifetime to alter the
late evolution of the cosmos.  We will see below that this
assumption can be justified given our results.
In order to proceed, we need to solve \eq{maTi} for $T_i$ and
hence we will need an expression for $m_a(T_i)$.

Following the treatment in Ref.~\cite{Preskill:1982cy}, $m_a(T_i)$ can
be obtained from
\beq
m_a^2(T) = \left(\frac{1}{f_a^2}\right) \left[\frac{\partial^2 F(T, \Theta)}
{\partial \Theta^2}\right]_{{\bar \Theta} = 0},
\label{ma2T}
\eeq
where $\Theta$ is the ``strong" CP violating angle and
$F(T, \Theta)$ is the free energy density in
the technicolor model.  We note that $\Theta$ is assumed
to be physically non-zero in the absence of a PQ mechanism.
Therefore, we require all chiral
symmetries to be broken at the weak scale.
This will ensure that there are no massless PGB's left in the
low energy spectrum, which is a desired
feature in any realistic model.

We have \cite{Gross:1980br}
\beq
\left[\frac{\partial^2 F(T, \Theta)}
{\partial \Theta^2}\right]_{{\bar \Theta} = 0} \simeq
\int_0^\infty d\rho\, n(\rho, T = 0),
\label{F}
\eeq
where instanton density $n(\rho, T=0)$,
as a function of instanton size
$\rho$, is given by \cite{Gross:1980br}:
\beq
n(\rho, 0) = \frac{C_{N_T}}{\rho^5}
\left(\frac{4 \pi^2}{g^2}\right)^{2 N_T}
\left(\prod_{i=1}^{2 N_D} \xi \rho m_i \right)
e^{- 8 \pi^2/g^2},
\label{n}
\eeq
with
$
4 \pi^2/g^2 = (1/6)(11 N_T - 4 N_D)\,
\ln[1/(\rho \LT)],
$
$
C_{N_T} \simeq 0.26\,
\xi^{-(N_T-2)} \, [(N_T - 1)! (N_T - 2)!]^{-1},
$
and $\xi \simeq 1.34$; $m_i$ are the techni-quark masses.

We have used the high temperature dilute instanton gas
approximation in writing \eq{F} \cite{Preskill:1982cy}, corresponding
to a sum of $T=0$ solutions.  However, to regulate
the infrared divergence of the integral, we use
the thermal cutoff $\rho_c = (\pi T)^{-1}$ \cite{Gross:1980br}.
Hence, we have
\begin{eqnarray}
\nonumber
m_a(T)&\simeq& \xi^{N_D} \kappa^{N_T} \left(\frac{C_{N_T}}
{2\eta}\right)^{1/2}
\left(\frac{m_Q}{\LT}\right)^{N_D} \\
&\times&\left(\frac{\LT}{\pi T}\right)^{\eta}
\left(\frac{\LT^2}{f_a}\right) \ln^{N_T}(\pi T/\LT),
\label{maT}
\end{eqnarray}
where $\kappa \equiv (1/6)(11 N_T - 4 N_D)$,
$\eta \equiv  \kappa + N_D - 2$,
and for simplicity $m_i = m_Q$, $\forall i$.
The quantity $m_Q$ enters determinants of Dirac operators
\cite{'tHooft:1976fv} involved in deriving \eq{n}.
By analogy with the relation between pion and quark masses in
QCD, we can write
\beq
m_Q = \frac{m_0^2 F_T^2}{2 \LT^3}
.
\label{mQ}
\eeq
\eq{mQ} gives the `hard masses'
of techniquarks \cite{Lane:1989ej,Lane:1991qh}.
Using \eq{maT}, we can write \eq{maTi} in the following form
\beq
\lambda \, (m_Q/\LT)^{N_D} \omega^{-(\eta+2)}
\ln^{N_T}(\omega) \simeq f_a/M_P,
\label{maTiII}
\eeq
where $\lambda \equiv (\pi^2/5)(2 \, g_{si}\, \eta)^{-1/2}
\, \xi^{N_D} \, \kappa^{N_T}\,
C_{N_T}^{1/2}$ and $\omega \equiv \pi T_i/\LT$.

The present energy density of the axions
$\Omega_a  \equiv \rho_a/\rho_c$,
where $\rho_c \simeq (2.6 \times 10^{-3} \, {\rm eV})^4$
is the critical energy density of the
universe \cite{Eidelman:2004wy}, is given by
\beq
\Omega_a \simeq
\left(\frac{10^{11} g_{sf}}
{N_D \, g_{si}^{1/2}}\right)\!\!
\left(\frac{A_i}{f_a}\right)^2\!\!\!
\left(\frac{m_0}{10^2~\!{\rm GeV}}\right)\!\!
\left(\frac{v_0}{\omega\LT}\right)\!\!
\left(\frac{f_a}{M_P}\right),
\label{Omegaa}
\eeq
with $T_f \simeq
2.3 \times 10^{-4}$~eV, the present temperature of
cosmic microwave background \cite{Eidelman:2004wy},
and $m_{af}$ given by \eq{maTC}.

We assume $g_{si}$ to be given by the degrees of freedom
contained in SM$-\{H\}$ plus technicolor, as a
function of $(N_T, N_D)$
\beq
g_{si} \simeq
102.75 + 7 N_T N_D + 2(N_T^2 - 1).
\label{gsi}
\eeq
Here, the value of $g_{sf}$ is set by the degrees of freedom at late
times, after $e^+ e^-$ annihilation;
$g_{sf} \simeq 4.13$.

We can now use the above results to study the
effects of axions on cosmology.  We will assume $m_0 = 100$~GeV,
hereafter.  Then, given
$(N_T, N_D)$ and $A_i/f_a$, \eq{maTiII} implies that $\Omega_a$ is
only a function of $\omega$.  Thus, bounds on $\Omega_a$ translate into
bounds on $f_a/M_P$, via \eq{maTiII}.

\vskip 0.5cm
\begin{table}
[t]
\begin{tabular}{lccc}
\hline \hline
$(N_T, N_D)$
 & $f_a/M_P\,<\;\;$
 &   $m_a \;{\rm (keV)}\,>\;\;$
 &   $\tau_a \;{\rm (yr)}\,<\;\;$ \\
\hline
$(4, 1)$
 &  $1.1 \times 10^{-9}$
 &  $0.94$
 &  $1.6\times10^{14}$ \\
$(4, 2)$
 &  $8.2 \times 10^{-10}$
 &  $0.62$
 &  $3.2\times 10^{14}$ \\
$(6, 2)$
 &   $6.5\times10^{-10}$
 &   $0.79$
 &   $9.4\times 10^{13}$\\
\hline\hline
\end{tabular}
\caption{Overclosure upper bounds on $f_a/M_P$, as a function of
technicolor parameters $(N_T, N_D)$.  The corresponding
lower bounds on $m_a$ and upper bounds on $\tau_a$ are also listed.
For these results, $A_i/f_a = 1$ and $m_0 = 100$~GeV.
}
\label{t1}
\end{table}

Avoiding the overclosure of the universe requires $\Omega_a < 1$.
For some choices of technicolor model parameters, the
overclosure upper bounds on $f_a/M_P$ are given in
Table~\ref{t1}, where we have set $A_i/f_a = 1$.
The first set of parameters is
motivated by minimality \cite{Hill:2002ap} and the other
sets show the effects of modest variations in $N_D$ and $N_T$.

In Table~\ref{t1}, we also present
$m_a$, from \eq{maTC}, and the lifetime $\tau_a$ of the axion
for maximal $f_a/M_P$.
Here, $\tau_a$ is given by the width
$\Gamma(a \to \gamma \gamma)$ for decay into photons \cite{GR}
\beq \tau_a^{-1} = \Gamma(a \to \gamma \gamma) =
\frac{\alpha^2\, m_a^3} {64 \pi^3\, f_a^2},
\label{taua}
\eeq
where, $\alpha \simeq 1/137$ is the fine structure constant.
We see that, in all cases considered in Table~\ref{t1},
$\tau_a$ is much larger than the age of the universe
($\sim 10^{10}$~yr), justifying the implementation of the
overclosure bound at late times.

Given our results, an interesting possibility
is that the axions of Table~\ref{t1}, with values
of $f_a/M_P$ just below the upper bound,
can be good dark matter candidates provided that
$\Omega_a \simeq 0.2$ \cite{Spergel:2006hy}.  Note that
sub-critical values of $f_a/M_P$ imply shorter lifetimes
than listed in Table~\ref{t1}.
The lifetime $\tau_x$ of a particle, with
mass $m_x \sim 1$~keV and relic density
$\Omega_x = \rho_x/\rho_c$,
is constrained by diffuse x-ray background
data \cite{Kawasaki:1997ah}:
\beq
\tau_x \gsim 3.3 \times 10^{19}\, \Omega_x 
\left(\frac{m_x}{{\rm~keV}}\right)^{-0.6} \!{\rm~yr}.
\label{x-ray}
\eeq
The bound in (\ref{x-ray}) can be satisfied for
$m_a \sim 1$~keV if dark matter axions have
$\tau_a \gsim 10^{19}$~yr.  This means that the cases
presented in Table~\ref{t1} cannot be dark matter.
Nonetheless, a somewhat different choice of initial conditions
can change these conclusions.

\vskip 0.5cm
\begin{table}
[h]
\begin{tabular}{lcccc}
\hline \hline
$(N_T, N_D)$
 & $f_a/M_P$
 &   $m_a \;{\rm (keV)}$
 &   $\tau_a \;{\rm (yr)}$
 &   $\tau_x \;{\rm (yr)}$\\
\hline
$(4, 1)$
 &  $1.4 \times 10^{-8}$
 &  $0.071$
 &  $6.3\times10^{19}$
 &  $3.2\times10^{19}$ \\
$(4, 2)$
 &  $1.1 \times 10^{-8}$
 &  $0.048$
 &  $1.2\times10^{20}$
 &  $4.1\times 10^{19}$ \\
$(6, 2)$
 &   $9.4\times10^{-9}$
 &   $0.054$
 &  $6.2\times10^{19}$
 &   $3.8\times 10^{19}$\\
\hline\hline
\end{tabular}
\caption{Axion parameters resulting in dark matter density
$\Omega_a = 0.2$.  For these results, $A_i/f_a = 0.1$
and $m_0 = 100$~GeV.
}
\label{t2}
\end{table}

The results in Table~\ref{t1}
have been obtained with $A_i/f_a = 1$.
However, $A_i/f_a = 0.1$ is also a plausible choice.
Using this value for the initial
amplitude, we have listed the
parameters of axions with dark matter energy
density $\Omega_a = 0.2$, in Table~\ref{t2}.
The results show that these axions can satisfy the
x-ray background bound (\ref{x-ray}) and be good
dark matter candidates, within a typical range of parameters.
Note that the parameters of the above dark matter
axions are very different from those associated with
the strong CP problem, for $A_i/f_a \sim 0.1$:
$f_a^{(QCD)}/M_P \sim 10^{-5}$ and $m_a^{(QCD)} \sim 10^{-8}$~eV
\cite{Preskill:1982cy}.  The discovery of the
technicolor related axions we have studied may therefore pose an
experimental challenge.

Various astrophysical constraints, such as supernova over-cooling,
place a lower bound on $f_a/M_P$, where the precise bound has
model-dependence.   However, we may take $f_a/M_P \gsim 8\times
10^{-11}$ for light axion-like particles \cite{Eidelman:2004wy}.
In principle, values of $f_a/M_P$ larger than this lower bound
but below the value resulting in dark matter
density $\Omega_a \simeq 0.2$ could be allowed.
However, the axion lifetime $\tau_a \sim f_a^5$, and
the bound in (\ref{x-ray}) is generally a stronger constraint
than astrophysical considerations.

Nonetheless, for small enough relic densities,
values of $f_a/M_P$ near the astrophysical lower bound
could still be allowed.  For example, if $A_i/f_a \sim 10^{-2}$,
the $(4, 1)$ case with $f_a/M_P \simeq 8\times 10^{-11}$ gives
$m_a \simeq 12.8$~keV, $\tau_a \simeq 3.3 \times 10^8$~yr,
and $\Omega_a \sim 5\times 10^{-6}$.  This set of values
is roughly consistent with the x-ray background bounds and may
have an observable impact on the early reionization history of
the universe \cite{Kasuya:2003sm,Kasuya:2004qk}.  Definitive
bounds and predictions for the effect of such axions on
early reionization require specific models and
a more detailed study and is
hence outside the scope of this work.

Given the high scales involved in PQ symmetry breaking, it is
interesting to ask whether such scales can be important for other
reasons in Nature.  Generation of small neutrino masses typically
requires operators suppressed by high scales and is thus a good
candidate.  Assuming that neutrinos are Dirac particles, the operator
\beq
O_\nu \sim \frac{({\bar Q_L}Q_R) (L \nu_R)}{f_a^2},
\label{Onu}
\eeq
generates a small neutrino mass $m_\nu$ upon EWSB via the
techni-condensate $\vev{{\bar Q_L}Q_R} \approx \LT^3$.
In \eq{Onu}, $L$ is the SM lepton doublet and $\nu_R$ is
a right handed neutrino; the coefficient of $O_\nu$ is assumed
to be ${\cal O}(1)$.  For example, the $(4, 1)$ case,
with $f_a$ near the astrophysical lower bound,
yields $m_\nu \approx 0.1$~eV,
which can accommodate the neutrino oscillation data.
As discussed before, this set of parameters can be allowed for
$A_i/f_a \lsim 10^{-2}$ and may lead to observable early reionization
effects.

In summary, we have studied the possibility that technicolor dynamics
is the source of mass for axion-like particles.  This assumes a techni-anomalous
PQ symmetry.  Given existing experimental constraints and
the potentially important role of high scales in Nature,
the PQ scale may be large; $f_a \gsim 10^{10}$~GeV.
Hence, the axions we have studied are light,
with $m_a \lsim 0.1$~keV, and `invisible'.  Assuming
a generic parametrization of technicolor models based on
the size of their gauge groups and number of weak doublets,
we derived overclosure bounds on the scale $f_a$.
However, bounds from diffuse x-ray background data further constrain 
$f_a$.  We showed that for typical
sets of parameters, these axions can be suitable
dark matter and consistent with the x-ray data, if the initial amplitude of axion primordial oscillations is mildly suppressed.  
Detection of this type of dark matter could be
difficult.   For smaller values
of $f_a$, the axion decay into photons
may impact early reionization history of the universe.
Such values of $f_a$ can lead to acceptable
neutrino masses through a seesaw
involving the technicolor condensate.

\acknowledgments

We would like to thank S. Chivukula, P. Huber, 
H. Murayama, F. Petriello, and P. Sikivie for discussions, and D. Chung 
for comments on a preliminary manuscript.
This work was supported in part by the United States
Department of Energy under Grant Contract No. DE-FG02-95ER40896 and
by the P.A.M. Dirac Fellowship, awarded by the
Department of Physics at the University of Wisconsin-Madison.



\begin{thebibliography}{99}

\bibitem{Weinberg:1975gm}
  S.~Weinberg,
  Phys.\ Rev.\ D {\bf 13}, 974 (1976);
  L.~Susskind,
  Phys.\ Rev.\ D {\bf 20}, 2619 (1979).

\bibitem{Peccei:1977hh}
  R.~D.~Peccei and H.~R.~Quinn,
  Phys.\ Rev.\ Lett.\  {\bf 38}, 1440 (1977);
  Phys.\ Rev.\ D {\bf 16}, 1791 (1977).

\bibitem{Weinberg:1977ma}
  S.~Weinberg,
  Phys.\ Rev.\ Lett.\  {\bf 40}, 223 (1978);
  F.~Wilczek,
  {\it ibid.}  {\bf 40}, 279 (1978).

\bibitem{Eidelman:2004wy}
  S.~Eidelman {\it et al.}  [Particle Data Group],
  Phys.\ Lett.\ B {\bf 592}, 1 (2004).

\bibitem{Preskill:1982cy}
  J.~Preskill, M.~B.~Wise and F.~Wilczek,
  Phys.\ Lett.\ B {\bf 120}, 127 (1983).

\bibitem{Abbott:1982af}
  L.~F.~Abbott and P.~Sikivie,
  Phys.\ Lett.\ B {\bf 120}, 133 (1983).

\bibitem{Dine:1982ah}
  M.~Dine and W.~Fischler,
  Phys.\ Lett.\ B {\bf 120}, 137 (1983).


\bibitem{TCcosmo}  For other possible effects of technicolor dynamics
on cosmology, see, for example:
  P.~H.~Frampton and S.~L.~Glashow,
  %
  Phys.\ Rev.\ Lett.\  {\bf 44}, 1481 (1980);
  S.~Nussinov,
  %
  Phys.\ Lett.\ B {\bf 165}, 55 (1985);
  S.~Dodelson,
  %
  Phys.\ Rev.\ D {\bf 40}, 3252 (1989);
R.~S.~Chivukula and T.~P.~Walker,
  %
  Nucl.\ Phys.\ B {\bf 329}, 445 (1990);
J.~A.~Frieman and G.~F.~Giudice,
  %
  Nucl.\ Phys.\ B {\bf 355}, 162 (1991);
J.~Bagnasco, M.~Dine and S.~D.~Thomas,
  %
  Phys.\ Lett.\ B {\bf 320}, 99 (1994)
  [arXiv:hep-ph/9310290].

\bibitem{Sikivie:1982qv}
  P.~Sikivie,
  Phys.\ Rev.\ Lett.\  {\bf 48}, 1156 (1982).

\bibitem{PS}

We thank P. Sikivie for comments on these issues.


\bibitem{GM}
  M.~Dine and A.~E.~Nelson,
  Phys.\ Rev.\ D {\bf 48}, 1277 (1993)
  [arXiv:hep-ph/9303230];
  M.~Dine, A.~E.~Nelson and Y.~Shirman,
  Phys.\ Rev.\ D {\bf 51}, 1362 (1995)
  [arXiv:hep-ph/9408384]; 
  M.~Dine, A.~E.~Nelson, Y.~Nir and Y.~Shirman,
  Phys.\ Rev.\ D {\bf 53}, 2658 (1996)
  [arXiv:hep-ph/9507378].


\bibitem{Hill:2002ap}
  C.~T.~Hill and E.~H.~Simmons,
  Phys.\ Rept.\  {\bf 381}, 235 (2003)
  [Erratum-ibid.\  {\bf 390}, 553 (2004)]
  [arXiv:hep-ph/0203079].

\bibitem{Georgi:1986df}
  H.~Georgi, D.~B.~Kaplan and L.~Randall,
  Phys.\ Lett.\ B {\bf 169}, 73 (1986).

\bibitem{Gross:1980br}
  D.~J.~Gross, R.~D.~Pisarski and L.~G.~Yaffe,
  Rev.\ Mod.\ Phys.\  {\bf 53}, 43 (1981).

\bibitem{'tHooft:1976fv}
  G.~'t Hooft,
  Phys.\ Rev.\ D {\bf 14}, 3432 (1976)
  [Erratum-ibid.\ D {\bf 18}, 2199 (1978)].


\bibitem{Lane:1989ej}
  K.~D.~Lane and E.~Eichten,
  %
  Phys.\ Lett.\ B {\bf 222}, 274 (1989).

\bibitem{Lane:1991qh}
  K.~D.~Lane and M.~V.~Ramana,
  %
  Phys.\ Rev.\ D {\bf 44}, 2678 (1991).

\bibitem{GR}
G.~G.~Raffelt, {\it Stars as Laboratories for Fundamental Physics}
(The University of Chicago Press, 1996).

\bibitem{Spergel:2006hy}
  D.~N.~Spergel {\it et al.},
  arXiv:astro-ph/0603449.

\bibitem{Kawasaki:1997ah}
  M.~Kawasaki and T.~Yanagida,
  Phys.\ Lett.\ B {\bf 399}, 45 (1997)
  [arXiv:hep-ph/9701346].

\bibitem{Kasuya:2003sm}
  S.~Kasuya, M.~Kawasaki and N.~Sugiyama,
  Phys.\ Rev.\ D {\bf 69}, 023512 (2004)
  [arXiv:astro-ph/0309434];

\bibitem{Kasuya:2004qk}
  S.~Kasuya and M.~Kawasaki,
  Phys.\ Rev.\ D {\bf 70}, 103519 (2004)
  [arXiv:astro-ph/0409419].















\end{thebibliography}
\end{document}